\documentclass[conference]{IEEEtran}

\newtheorem{theorem}{Theorem}
\newtheorem{lemma}{Lemma}
\newtheorem{corollary}{Corollary}
\newtheorem{definition}{Definition}

\usepackage[cmex10]{amsmath}
\usepackage{amssymb}

\usepackage[dvips]{graphicx}

\usepackage{psfrag}

\IEEEoverridecommandlockouts

\begin{document}
\title{On the Sum Capacity of $K$-user Cascade Gaussian Z-Interference Channel}

\author{\IEEEauthorblockN{Yuanpeng Liu, Elza Erkip}
\IEEEauthorblockA{ECE Department, Polytechnic Institute of New York University\\
yliu20@students.poly.edu, elza@poly.edu}
\thanks{This work is supported in part by NSF grant No. 0635177.}}

\maketitle


\begin{abstract}
A $K$-user cascade Gaussian Z-interference channel is a subclass of the general $K$-user Gaussian interference channel, where each user, except the first one, experiences interference only from the previous user. Under simple Han-Kobayashi schemes assuming Gaussian inputs and no time sharing, it is shown that the maximum sum rate is achieved by each user transmitting either common or private signals. For $K=3$, channel conditions under which the achieved sum rate is either equal to or within 0.5 bits to the sum capacity are identified.

\end{abstract}

\section{Introduction}
Gaussian interference channel (GIC) models a communication scenario where the received signal is not only subject to Gaussian noise but also the interference coming from other transmissions. Despite its simple form, the capacity region of the 2-user GIC is unknown except for some special cases such as the strong interference regime \cite{Sato}. Additionally the sum capacity in the noisy regime was derived concurrently in \cite{Anna_Veer}\cite{Mota_Khan}\cite{Shang_Kranmer}. For a general 2-user GIC the best achievable rate region is given by the Han-Kobayashi (HK) scheme \cite{HK} where the information is split into common and private components. However the computation of the general HK rate region is very difficult due to a large degrees of freedom involved and the lack of knowledge of the optimal input distribution. Nevertheless in \cite{Etkin} Etkin, Tse and Wang showed that a Gaussian-input HK scheme without time sharing, where the power of the private information is set to the noise level, suffices to achieve the capacity region of a 2-user GIC to within one bit. Notably all the aforementioned capacity results reside on the class of simple HK schemes (Gaussian inputs, no time sharing) with different power splitting. When one is interested in a $K$-user GIC ($K>2$), it is tempting to consider the generalization of the simple HK scheme in a $K$-user setup. Unfortunately the simple HK scheme is generally insufficient. As shown in \cite{Bresler} it does not achieve the degree of freedom of a many-to-one GIC, a special case of a general $K$-user GIC. In this paper, we seek to find a subclass of a general $K$-user GIC where a simple HK scheme achieves the sum capacity.

\begin{figure}[htb]
    \centering
    \includegraphics[width=80mm]{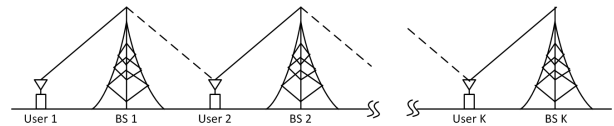}
    \caption{In this particular Wyner cellular network, user $i$ is associated with base station $i$. Only the interference coming from the nearest non-intended base station is considered. Interference links are denoted by dashed lines.}
    \label{wynercell}
\end{figure}

Motivated by the fact that the sum capacity of a 2-user one-sided GIC (here we refer it as Gaussian Z-interference channel or GZIC) is known for all channel conditions \cite{Sason}, we consider the generalization of the channel model to the $K$-user case. Each user, except for the first interference-free user, experiences only one interference component coming from the previous user. Hence every two adjacent users form a GZIC and all the GZICs connect in a cascade fashion shown in Fig. \ref{chnfig} (in Section II). We refer this channel as cascade Gaussian Z-interference channel (CGZIC). For a physical interpretation, consider a Wyner cellular network \cite{Wyner} where the base stations are linearly aligned and each base station serves one user, which is located at the edge of the cell as shown in Fig. \ref{wynercell}. The $K$-user CGZIC models the downlink channel of such network.

In this paper, we first obtain the sum rate optimal power splitting within a class of simple HK schemes assuming Gaussian inputs and no time sharing. We show that there is no need for information splitting and each user transmits either common or private information depending on a recursively-defined channel condition. Then for a 3-user CGZIC, by providing upper-bounds, we present the sum capacity results in various regimes. For a modified noisy regime, a class of mixed regimes and the strong interferece regime, the exact sum capacity is derived. For a different class of mixed regimes, the sum capacity to within 0.5 bits is derived. All of the achievable schemes are based on the above optimized simple HK scheme. The generalization to the $K$-user case ($K>3$) is also discussed.

In related work \cite{Zhou}, Zhou and Yu considered a similar channel model named cyclic GIC where the Etkin-Tse-Wang scheme was shown to achieve the capacity region to within some constant number of bits in the weak interference regime. The CGZIC considered in this paper is a special case of a cyclic GIC, where one interference link is removed. In return, the simplified model allows us to obtain stronger results in terms of rate sum. In particular, the optimized simple HK scheme proposed in this paper achieves a higher sum rate than the Etkin-Tse-Wang scheme when applied to the CGZIC. Furthermore we are able to obtain the exact sum capacity in some regimes, as opposed to the approximation approach taken in \cite{Zhou}. Also the simplified channel allows us to obtain capacity results in a broader range of channel conditions.

This paper is organized as the follows. Section II formally defines the channel model. For a class of simple HK schemes, the maximum sum rate is derived in Section III. Section IV presents the sum capacity results for various regimes and Section V concludes the paper by illustration and discussion.

\emph{Notation}: $\mathcal{C}(x)\triangleq\frac{1}{2}\textrm{log}_2(1+x)$, $\mathcal{C}^{-1}(y)\triangleq2^{2y}-1$. $X^n$ denotes $(X_1,...,X_n)$. Whenever we write $Y_G$, the subscript G is used to denote that the distribution is Gaussian. $h(\cdot)$ denotes differential entropy. The logarithms are in base 2.

\section{Channel Model}
\begin{figure}[htb]
    \centering
    \includegraphics[width=43mm]{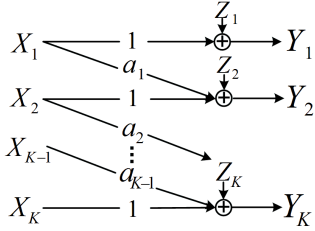}
    \caption{$K$-user cascade Gaussian Z-interference channel}
    \label{chnfig}
\end{figure}

We consider a general $K$-user CGZIC with arbitrary channel coefficients and noise variances, which can be transformed into the standard form shown in Fig. \ref{chnfig},
\begin{equation*}
    \begin{split}
        \label{channelmodel}
        Y_1&=X_1+Z_1\\
        Y_i&=X_i + a_{i-1}X_{i-1}+Z_i, \quad i=2,3,...,K,
    \end{split}
\end{equation*}
where $Z_i$ is the i.i.d. real Gaussian noise with zero mean and unit variance, i.e. $Z_i\sim \mathcal{N}(0,1)$. Here $X_i$ and $Y_i$ denote the input and output of user $i$ respectively. We assume the channel input is subject to an average power constraint $P_i$.

For a given rate tuple $(R_1,...,R_K)$ and number of channel uses $n$, user $i$ maps a message $W_i,W_i\in\{1,...,2^{nR_i}\}$ into a codeword, $X_i^n$, via an encoding function such that the power constraint is satisfied. Receiver $i$ produces an estimate $\hat{W}_i$ of message $W_i$ based on the channel output $Y_i^n$ via a decoding function. Rate tuple $(R_1,...,R_K)$ is said to be achievable if there exits encoding and decoding functions such that for all $i\in \{1,...,K\}$, the average error probability of user $i$ $P_{e,i}=E(P_{r}(\hat{W}_i \neq W_i))$, where the expectation is over all messages, approaches zero as $n$ approaches infinity. The capacity region is defined as the union of all achievable rate tuples.

\section{Sum Rate of A Class of Simple HK Schemes}
We define a class of simple HK schemes as the follows:
\begin{definition}
 In a {\em simple HK scheme}, user $i$, $i=1,...,K$, employs a Gaussian codebook and splits its power into two parts, the common $\gamma_i P_i$ and the private $(1-\gamma_i)P_i$, $0\leq\gamma_i\leq 1$. Receiver $i$ decodes the common part of the interference and its own signal jointly by treating the private part as noise. Hence each scheme in the class is uniquely specified by the power splitting tuple $(\gamma_1,...,\gamma_K)$.
\end{definition}

Note that the simple HK scheme, of which the rate region for a given channel is solely determined by the power splitting tuple, is more tractable than the general HK scheme, where besides the power splitting, the characterization of the rate region also depends on input distributions and time-sharing. Previous works summarized in the introduction have focused on tuning the power splitting of the simple HK scheme in such a way that the sum rate coincides (within a constant gap) with the upper-bound and therefore capacity results are obtained. In this section we maximize the sum rate of the simple HK scheme for a $K$-user CGZIC. In the following section, we provide upper-bounds and establish capacity results.

\begin{theorem}
\label{sumrate}
The maximum sum rate of a $K$-user CGZIC under a class of simple HK schemes is given by
\begin{equation}
\label{sumrateexp}
    R_{\textrm{sum}}=\sum_{i=1}^K\mathcal{C}(h_i^2P_i),
\end{equation}
where $h_1=1$ and for $i=2,...,K$,
\begin{equation}
\label{effchnl}
    h_i=\begin{cases}
        \mbox{\fontsize{9}{10.8}$\sqrt{\frac{1}{1+a_{i-1}^2P_{i-1}}},  \ a_{i-1}\leq h_{i-1}$}\\
        \mbox{\fontsize{9}{10.8}$\min\left\{ \sqrt{\frac{(a_{i-1}^2-h_{i-1}^2)P_{i-1}+P_i}{P_i+h_{i-1}^2P_{i-1}P_i}}, 1 \right\}, \ a_{i-1}> h_{i-1}$}
    \end{cases}.
\end{equation}
The optimal power splitting tuple is given by $(\gamma_1^*,\gamma_2^*,...,\gamma_K^*)$, where $\gamma_K^*$ is an arbitrary real number in the interval $[0,1]$ and
\begin{equation}
    \label{splitting}
    \gamma_i^*=\begin{cases}
        0,\quad &\quad a_{i}\leq h_{i}\\
        1,\quad &\quad a_{i}>h_{i}
    \end{cases}\quad \textrm{for } i=1,...,K-1.
\end{equation}
\end{theorem}

\begin{IEEEproof}
Suppose $\mathbf{R}^*=(R_1^*,...,R_K^*)$ is a maximum sum rate achieving rate vector under a class of simple HK schemes. Let $\mathbf{\Gamma}^*=(\gamma_1^*,\gamma_2^*,...,\gamma_K^* )$ be the corresponding power splitting tuple. Now consider a new power splitting tuple $\mathbf{\Gamma}'=(\gamma_1,\gamma_2^*,...,\gamma_K^* )$ where
\begin{equation*}
    \gamma_1=\begin{cases}
    0,\quad &\quad a_1\leq h_1\\
    1,\quad &\quad a_1> h_1
    \end{cases}.
\end{equation*}
Note that $\mathbf{\Gamma}'$ differs from $\mathbf{\Gamma}^*$ only by the first element. Let $\mathbf{R}'=(R_1',...,R_K')$ denote the rate vector associated with $\mathbf{\Gamma}'$ scheme. Next we will show $\sum_{i=1}^K R_i'$ is at least as large as $\sum_{i=1}^K R_i^*$ and hence $\mathbf{\Gamma}'$ is also sum rate optimal. Then we can proceed repeatedly in a similarly fashion until all $\gamma_i^*$ is updated according to (\ref{splitting}). Now let $r_1^*=\mathcal{C}(h_1^2P_1)=\mathcal{C}(P_1)$ and for $i=2,...,K$,
\begin{align}
\label{ristar}
    &r_i^*=\mathcal{C}(h_i^2P_i)\notag\\
    &=\begin{cases}
    \mbox{\fontsize{8}{9.6}$\mathcal{C}(\frac{P_{i}}{1+a_{i-1}^2P_{i-1}}),\quad a_{i-1} \leq h_{i-1}$}\\
    \mbox{\fontsize{8}{9.6}$\min\{\mathcal{C}(a_{i-1}^2P_{i-1}+P_{i})-\mathcal{C}(h_{i-1}^2P_{i-1}),\mathcal{C}(P_{i})\},\ a_{i-1} > h_{i-1}$}
    \end{cases}
\end{align}
where $h_i$ is defined in (\ref{effchnl}).

Case 1. If $R_2^*< r_2^*$, we argue that $\mathbf{R}'=(R_1',...,R_K')$ is achievable under $\mathbf{\Gamma}'$, where
    \begin{align*}
        R_1'=r_1^*=\mathcal{C}(h_1^2P_1),\quad R_i'=R_i^* \textrm{ for } i=2,...,K.
    \end{align*}
    We need to show that with $\mathbf{\Gamma}'$ and rates $\mathbf{R}'$, receiver $i$ can reliably decode the signal sent from user $i$ for all $i=1,...,K$. Note that rate vector $(R_1',r_2^*)$ gives the sum capacity of a 2-user GZIC when the power splitting $\gamma_1$ is employed by user 1 \cite{Sason}. Since $R_2'=R_2^*< r_2^*$, rate vector $(R_1',R_2')$ lies in the capacity region of the 2-user GZIC formed by user 1 and 2. Hence decoding at receiver 1 and 2 will be successful. Recall that $\mathbf{\Gamma}'$ and $\mathbf{R}'$ differ from $\mathbf{\Gamma}^*$ and $\mathbf{R}^*$ only by the first element respectively. Under power splitting $\mathbf{\Gamma}'$, receiver $i$ can reliably decode at the rate $R_i'=R_i^*$ for $i=3,...,K$, since we assume $\mathbf{R}^*$ is achievable with scheme $\mathbf{\Gamma}^*$. By assumption $\sum_{i=1}^K R_i' \leq \sum_{i=1}^K R_i^*$, we have $R_1'\leq R_1^*$. However, we also have $R_1' \geq R_1^*$ since $R_1'$ is the point to point capacity of user 1. Therefore $R_1^*=R_1'=r_1^*$.

Case 2. If $R_2^*\geq r_2^*$, we argue that $\mathbf{R}'=(R_1',...,R_K')$ is achievable under $\mathbf{\Gamma}'$ scheme, where
    \begin{align*}
        R_1'&=r_1^*=\mathcal{C}(h_1^2P_1),\quad R_2'=r_2^*,\\
        R_i'&=R_i^* \textrm{ for } i=3,...,K.
    \end{align*}
    The decodability at receiver $i$, $i=1,2,4,...,K$ can be shown using the above argument. Now we consider the decodability at receiver 3. For any $R_2^*\geq r_2^*$ and any decomposition $R_2^*=R_{2c}^*+R_{2p}^*$, where the subscript $c/p$ denotes common/private information, we can always find $R_{2c}'$ and $R_{2p}'$ such that $R_{2c}' \leq R_{2c}^*$, $R_{2p}' \leq R_{2p}^*$ and $R_{2c}'+R_{2p}'=R_2'=r_2^*$. If receiver 3 can decode the common interference and its own signal at rates $(R_{2c}^*,R_3^*)$, which is true by the assumption that $\mathbf{R}^*$ is achievable under $\mathbf{\Gamma}^*$, then rates $(R_{2c}', R_3^*)$ are also decodable at receiver 3 for the same power splitting tuple $(\gamma_2^*,\gamma_3^*)$ employed by user 2 and 3. Therefore rate vector $\mathbf{R}'=(R_1',...,R_K')$ is achievable under $\mathbf{\Gamma}'$ scheme. Again by assumption $\sum_{i=1}^K R_i' \leq \sum_{i=1}^K R_i^*$, we have $R_1'+R_2' \leq R_1^* + R_2^*$. However, it is clear that $R_1'+R_2'$ gives the sum capacity of the 2-user GZIC formed by user 1 and 2. Hence $R_1' + R_2' = R_1^* + R_2^*$, i.e. $\mathbf{\Gamma}'$ scheme is sum rate optimal. Note that the maximum sum rate achieving scheme may not be unique. Nevertheless, since we seek to find one such scheme, we let $R_1^*=R_1'=r_1^*$ and $R_2^*=R_2'=r_2^*$.

We have shown that there exists a maximum sum rate achieving scheme satisfying the following conditions:
\begin{align*}
    R_1^*&=r_1^*=\mathcal{C}(h_1^2P_1),\\
    R_2^*&\leq r_2^*,\\
    &=\begin{cases}
                    \mathcal{C}(\frac{P_2}{1+a_1^2P_1}), \ a_1 \leq h_1\\
                    \min\{\mathcal{C}(a_1^2P_1+P_2)-\mathcal{C}(h_1^2P_1),\mathcal{C}(P_2)\},\ a_1 > h_1
                    \end{cases},\\
    \gamma_1^*&=\begin{cases}
                0, \quad a_{1}\leq h_{1}\\
                1, \quad a_{1}>h_{1}
                \end{cases}.
\end{align*}
Since we have determined $R_1^*$, we can safely remove user 1 from the system and focus on the remaining users with an additional rate constraint on $R_2^*$. We can further drop this constraint by replacing user 2's direct link by $h_2$, $h_2=\sqrt{\frac{\mathcal{C}^{-1}(r_2^*)}{P_2}}$. Note that $R_2^*$ is upper-bounded by the point to point rate and $h_i\leq 1$, $i=1,...,K$. Now we are back to the problem for a $(K-1)$-user CGZIC starting with user 2 with a modified direct link. We can apply this procedure repeatedly until all users are considered. Notice that since user $K$ does not cause interference to any other users, there is no constraint on the power splitting user $K$ adopts.
\end{IEEEproof}

%


\section{Capacity Results}
In this section, we consider a 3-user CGZIC and present capacity results in various regimes. Extension to the $K$-user case is discussed in Section V. One trivial result is that if there exists a very strong interference link, say $a_1\geq \sqrt{1+P_2}$, this interference link could be removed without affecting the capacity region and hence the sum capacity of a 3-user CGZIC in this case reduces to the sum of a point to point AWGN channel capacity and the 2-user GZIC sum capacity. Without loss of generality, we assume no very strong interference link exists, i.e. $a_i<\sqrt{1+P_{i+1}}$ for $i=1,2$.


\subsection{Sum Capacity in the Noisy Interference Regime}
\begin{definition}
\label{defnoisy}
For a 3-user CGZIC, we say the interference is \textit{noisy} if the following condition is satisfied:
\begin{equation}
    \label{noisy}
    a_1^2+a_2^2(1+a_1^2P_1)^2\leq 1.
\end{equation}
\end{definition}
The following result is a direct application of Theorem 3 in \cite{S.K.C} to a 3-user CGZIC.
\begin{corollary}
\label{corollary}
The sum capacity of a 3-user CGZIC in the noisy interference regime is
\begin{equation*}
    C_{\textrm{sum}}=\mathcal{C}(P_1)+\mathcal{C}(\tfrac{P_2}{1+a_1^2P_1})+\mathcal{C}(\tfrac{P_3}{1+a_2^2P_2}).
\end{equation*}
\end{corollary}
\begin{IEEEproof}
In \cite{S.K.C}, the authors use $c_{ji}$ to denote the channel coefficient from the $j$th user to the $i$th receiver. Specifically for the 3-user CGZIC considered in this paper, $c_{13}=c_{21}=c_{31}=c_{32}=0$, $c_{12}=a_1^2$ and $c_{23}=a_2^2$. Substituting these values into equation (27)(28) in \cite{S.K.C}, we get the following channel condition under which treating interference as noise achieves sum capacity:
\begin{align*}
    a_1^2\leq(1-\rho_2^2)\rho_1^2,\quad a_2^2(1+a_1^2P_1)^2\leq (1-\rho_3^2)\rho_2^2,
\end{align*}
for some value $\rho_i\in[0,1]$, $i=1,2,3$. Such $\rho_i$ always exists if condition (\ref{noisy}) holds.
\end{IEEEproof}

Notice that condition (\ref{noisy}) implies $a_1\leq 1$ and $a_2 \leq \sqrt{\frac{1}{1+a_1^2P_1}}$. According to Theorem \ref{sumrate}, treating interference as noise at both receivers achieves the maximum sum rate of the simple HK scheme and Corollary \ref{corollary} shows that such sum rate is indeed the sum capacity.


\subsection{Capacity Region in the Strong Interference Regime}
\begin{definition}
\label{defstrong}
For a 3-user CGZIC, we say the interference is \textit{strong} if $1\leq a_1<\sqrt{1+P_2}$ and $1\leq a_2<\sqrt{1+P_3}$.
\end{definition}

\begin{theorem}
\label{strong}
The capacity region of a 3-user CGZIC in the strong interference regime is given by the set of all nonnegative rate triples $(R_1,R_2,R_3)$ satisfying the following conditions for $i=1,2,3$ and $j=1,2$:
\begin{align*}
    R_i&\leq\mathcal{C}(P_i),\\
    R_j&\leq\mathcal{C}(a_j^2P_j),\\
    R_j+R_{j+1}&\leq\mathcal{C}(a_j^2P_j+P_{j+1}).
\end{align*}
\end{theorem}
\begin{IEEEproof}
For the achievability, consider the simple HK scheme with power splitting tuple $(\gamma_1,\gamma_2,\gamma_3)=(1,1,\gamma_3)$, where $\gamma_3$ is arbitrary in the interval $[0,1]$. The converse follows the same argument in \cite{Sato}. Since $a_1\geq 1$ ($a_2\geq 1$), receiver 2 (3) will be able to decode the interference signal if its own signal is decodable and hence can be removed. Therefore the rates of user 1 (2) and 2 (3) should lie in the capacity region of the multi-access channel at receiver 2 (3) in Theorem \ref{strong}.
\end{IEEEproof}

Using Fourier-Motzkin elimination, we have the following.
\begin{corollary}
\label{strongcoro}
The sum capacity of a 3-user CGZIC in the strong interference regime is
\begin{align*}
    C_{\textrm{sum}}=\min\left\{\mathcal{C}(P_1)+\mathcal{C}(a_2^2P_2+P_3), \mathcal{C}(P_3)+\mathcal{C}(a_1^2P_1+P_2)\right\}.
\end{align*}
\end{corollary}

In fact since Theorem \ref{strong} establishes the optimality of the simple HK scheme, the above sum capacity expression can be directly derived from Theorem \ref{sumrate} using the power splitting tuple $(\gamma_1,\gamma_2,\gamma_3)=(1,1,\gamma_3)$, where $\gamma_3$ is arbitrary in the interval $[0,1]$.


\subsection{Sum Capacity in the Mixed Interference Regimes}
\begin{definition}
For a 3-user CGZIC, we say the interference is in \textit{mixed regime I} if the following conditions hold: $a_1< 1$ and $\sqrt{\frac{1+P_3}{1+a_1^2P_1}}\leq a_2<\sqrt{1+P_3}$.
\end{definition}
\begin{theorem}
 The sum capacity of a 3-user CGZIC in the mixed regime I is
\begin{equation*}
    C_{\textrm{sum}}=\mathcal{C}(P_1)+\mathcal{C}(\tfrac{P_2}{1+a_1^2P_1})+\mathcal{C}(P_3).
\end{equation*}
\end{theorem}
\begin{IEEEproof}
For the achievability, consider the simple HK scheme with power splitting tuple $(\gamma_1,\gamma_2,\gamma_3)=(0,1,\gamma_3)$, where $\gamma_3$ is arbitrary in the interval $[0,1]$. According to Theorem \ref{sumrate}, the following sum rate is achievable:
\begin{align}
    R_{\textrm{sum}}&=\mathcal{C}(h_1^2P_1)+\mathcal{C}(h_2^2P_2)+\mathcal{C}(h_3^2P_3),\notag\\
    &=\mathcal{C}(P_1)+\mathcal{C}(\tfrac{P_2}{1+a_1^2P_1})+\mathcal{C}(h_3^2P_3),\label{mixedIeq1}\\
    &=\mathcal{C}(P_1)+\mathcal{C}(\tfrac{P_2}{1+a_1^2P_1})+\mathcal{C}(P_3)\label{mixedIeq2},
\end{align}
where (\ref{mixedIeq1}) is because $a_1<1$ and (\ref{mixedIeq2}) is due to $a_2\geq \sqrt{\frac{1+P_3}{1+a_1^2P_1}}$. The converse follows immediately by recognizing that the first two terms gives the sum capacity of the 2-user GZIC formed by user 1 and 2 in the weak interference regime ($a_1<1$).
\end{IEEEproof}

\begin{definition}
For a 3-user CGZIC, we say the interference is in \textit{mixed regime II} if the following conditions hold: $1\leq a_1<\sqrt{1+P_2}$ and $a_2\leq \sqrt{\frac{1}{1+a_1^2P_1}}$.
\end{definition}

\begin{theorem}
\label{mixedII}
The sum capacity of a 3-user CGZIC in the mixed regime II satisfies the following inequality:
\begin{equation*}
    C_{\textrm{sum}}\leq \mathcal{C}(a_1^2P_1+P_2) + \mathcal{C}(\tfrac{P_3}{1+a_2^2P_2}) + 0.5.
\end{equation*}
\end{theorem}
The proof of Theorem \ref{mixedII} relies on the following Lemma.
\begin{lemma}
\label{lemma}
Suppose $N_1,N_2,...,N_n$ are i.i.d. with an arbitrary distribution and $X_1,X_2,...,X_n$ $\sim$ i.i.d. $p(x)$ with $E_{p(x)}(X^2)\leq P$. Then we have
\begin{equation*}
    h(X^n+N^n)\leq h(X_G^n+N^n) + 0.5n,
\end{equation*}
where $X_G^n\sim$ i.i.d. $\mathcal{N}(0,P)$.
\end{lemma}
\begin{IEEEproof}
In \cite{Zamir}, the authors showed that Gaussian input incurs no more than 0.5-bit loss to the capacity of an additive arbitrarily distributed noise channel. Lemma \ref{lemma} is a direct consequence of the multi-letter version of their results, i.e.
\begin{equation*}
    I(X^n;X^n+N^n)\leq I(X_G^n;X_G^n+N^n)+0.5n.
\end{equation*}
The proof of the above inequality follows the exact same lines in \cite{Zamir} except that scalar random variables are now replaced by the corresponding vectors.
\end{IEEEproof}

We are now in position to prove Theorem \ref{mixedII}.
\begin{IEEEproof}[Proof of Theorem \ref{mixedII}]
For the achievability, consider the simple HK scheme with power splitting tuple $(\gamma_1,\gamma_2,\gamma_3)=(1,0,\gamma_3)$, where $\gamma_3$ is arbitrary in the interval $[0,1]$. Theorem \ref{sumrate} suggests optimality among simple HK schemes with the sum rate, $R_{sum}=\mathcal{C}(a_1^2P_1+P_2) + \mathcal{C}(\tfrac{P_3}{1+a_2^2P_2})$, being achievable.

The genie bounding approach \cite{Anna_Veer} combined with Lemma \ref{lemma} is used for the converse. If $a_1\geq 1$, $I(X_1;Y_2|X_2)\geq I(X_1;Y_1)$ for all input distributions since $Y_1$ is stochastically degraded with respect to $Y_2$ conditioned on $X_2$. As shown in \cite{Costa}, this implies the multi-letter version $I(X_1^n;Y_2^n|X_2^n)\geq I(X_1^n;Y_1^n)$. Let $S_2=X_2+a_1X_1+\eta_2N_2$ denote the signal given to receiver 2 by a genie, where $N_2\sim \mathcal{N}(0,1)$ is correlated with $Z_2$ with correlation coefficient $\rho_2$ and $\eta_2$ is some constant. For some $\epsilon_n$ such that $\lim_{n\rightarrow \infty}\epsilon_n=0$, we have
\begin{align}
    n&(R_1+R_2+R_3-\epsilon_n)\notag\\
    &\leq I(X_1^n;Y_1^n)+I(X_2^n;Y_2^n)+I(X_3^n;Y_3^n)\notag\\
    &\leq I(X_1^n,X_2^n;Y_2^n)+I(X_3^n;Y_3^n)\notag\\
    &\leq I(X_1^n,X_2^n;Y_2^n,S_2^n)+I(X_3^n;Y_3^n)\notag\\
    &=h(S_2^n)-h(S_2^n|X_1^n,X_2^n) + h(Y_2^n|S_2^n)\notag\\
    &\quad - h(Y_2^n|X_1^n,X_2^n,S_2^n) + h(Y_3^n)-h(Y_3^n|X_3^n)\notag\\
    &\leq h(S_2^n) - nh(S_{2G}|X_{1G},X_{2G})+h(Y_2^n|S_2^n)\notag\\
    &\quad - nh(Y_{2G}|X_{1G},X_{2G},S_{2G})+nh(Y_{3G})-h(Y_3^n|X_3^n)\label{mixedIIeq1}\\
    &\leq h(S_2^n) - nh(S_{2G}|X_{1G},X_{2G})+nh(Y_{2G}|S_{2G})\notag\\
    &\quad - nh(Y_{2G}|X_{1G},X_{2G},S_{2G})+nh(Y_{3G})-h(Y_3^n|X_3^n)\label{mixedIIeq2}
\end{align}
where (\ref{mixedIIeq1}) is due to the fact that the channel is memoryless, conditioning reduces entropy and given a covariance constraint, Gaussian distribution maximizes differential entropy and (\ref{mixedIIeq2}) is due to Lemma 1 in \cite{Anna_Veer}. Let $V$ denote a Gaussian r.v. independent of $X_2$ and $V_2,V_2\sim \mathcal{N}(0,a_2^2(a_1^2P_1+\eta_2^2))$, such that $Z_3=V_2+V$. Note that $V$ exists if $a_2^2(a_1^2P_1+\eta_2^2)\leq 1$. Then we have
\begin{align}
    h(&S_2^n)-h(Y_3^n|X_3^n)\notag\\
    &=h(X_2^n+a_1X_1^n+\eta_2N_2^n)-h(a_2X_2^n+Z_3^n)\notag\\
    &\leq h(a_2X_2^n+a_2a_1X_{1G}^n+a_2\eta_2N_2^n)-h(a_2X_2^n+Z_3^n)\notag\\
    &\quad+0.5n-n\log a_2\label{genie1}\\
    &= -I(V^n;a_2X_2^n+V_2^n+V^n)+0.5n-n\log a_2\label{genie2}\\
    &\leq -nI(V;a_2X_{2G}+V_2+V)+0.5n-n\log a_2\label{genie3}\\
    &=nh(S_{2G})-nh(Y_{3G}|X_{3G})+0.5n,\notag
\end{align}
where (\ref{genie1}) is due to Lemma \ref{lemma} and the identity $h(cX^n)=h(X^n)+n\log|c|$ for a constant $c$. Also $X_{1G}^n$ has i.i.d. entries. (\ref{genie2}) is because $V_2$ and $a_2(a_1X_{1G}+\eta_2N_2)$ have the same variance and are both independent of $X_2$. (\ref{genie3}) is due to the worst case noise result for the additive noise channel \cite{Diggavi}.

Due to Lemma 8 in \cite{Anna_Veer}, $I(X_{2G};S_{2G}|Y_{2G})=0$ and $I(X_{1G};S_{2G}|Y_{2G},X_{2G})=0$ iff $\eta_2\rho_2=1$. Therefore we have
\begin{align*}
    R_1+&R_2+R_3 -\epsilon_n \\
    &\leq I(X_{1G},X_{2G};Y_{2G},S_{2G})+I(X_{3G};Y_{3G})+0.5\\
    &=I(X_{1G},X_{2G};Y_{2G})+I(X_{2G};S_{2G}|Y_{2G})\\
    &\quad+I(X_{1G};S_{2G}|Y_{2G},X_{2G})+I(X_{3G};Y_{3G})+0.5\\
    &=\mathcal{C}(a_1^2P_1+P_2) + \mathcal{C}(\tfrac{P_3}{1+a_2^2P_2})+0.5,
\end{align*}
under the conditions: $a_2^2(a_1^2P_1+1/\rho_2^2)\leq 1$ for some $\rho_2\in[0,1]$. There always exists such $\rho_2$ if $a_2\leq \sqrt{\frac{1}{1+a_1^2P_1}}$.
\end{IEEEproof}

\section{Illustration and Discussion}
\begin{figure}[htb]
    \centering
    \includegraphics[width=50mm]{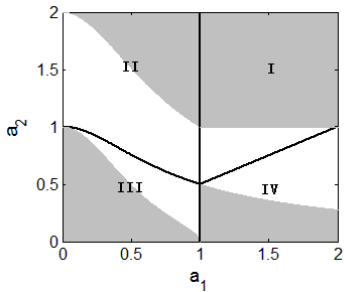}
    \caption{Four regimes of a 3-user CGZIC with $P_1=P_2=P_3=3$} 
    \label{numerfig}
\end{figure}

For a 3-user CGZIC, Fig. \ref{numerfig} shows four regimes for $(a_1,a_2)$. In regimes I, II, III and VI, user 1/ 2 transmitting common/common, private/common, private/private and common/private signals respectively maximizes the sum rate of a class of simple HK schemes using Theorem \ref{sumrate}. Specifically, the gray area in regime I corresponds to the strong interference regime where the sum rate (\ref{sumrateexp}) in Theorem \ref{sumrate} is the sum capacity. Likewise, the gray areas in regimes II, III and VI correspond to the mixed regime I, noisy regime and mixed regime II respectively where sum capacity results apply.

Regarding the generalization of the capacity results for a 3-user CGZIC to the $K$-user ($K>3$) case, similar to Definition \ref{defnoisy}, the channel condition for the $K$-user noisy interference regime can be identified using Theorem 3 \cite{S.K.C}, where the sum capacity is achieved by all receivers treating interference as noise. In the $K$-user strong interference regime, where all interference links are strong, i.e. $1\leq a_i\leq \sqrt{1+P_{i+1}}, i=1,...,K-1$, similar to Theorem \ref{strong}, it can be shown that only transmitting common information achieves the capacity region. We can also make use of the following Lemma to break the $K$-user CGZIC into smaller CGZICs.

\begin{lemma}
Consider a $K$-user CGZIC with $K>3$. Let $C_i^j$, $1\leq i,j\leq K$ and $i\leq j$, denote the sum capacity of a $(j-i+1)$-user CGZIC obtained by only considering users $\{i,...,j\}$. Assume there exists $k,k<K,$ such that $C_1^k$ is achieved by the simple HK scheme. If $a_k\geq \sqrt{1+P_{k+1}}h_k$ with $h_k$ defined in (\ref{effchnl}), then $C_1^K=C_1^k+C_{k+1}^K$.
\end{lemma}
\begin{IEEEproof}
By providing user $k+1$ the genie signal $X_k$, the interference link from user $k$ to user $k+1$ can be removed and the original channel can be decoupled into two parts, whose sum capacities are $C_1^k$ and $C_{k+1}^K$ respectively. Hence we have $C_1^K\leq C_1^k+C_{k+1}^K$. For the achievability, let the first $k$ users and the remaining $K-k$ users employ their respective sum capacity achieving schemes for $C_1^k$ and $C_{k+1}^K$. If $C_1^k$ is achieved by the simple HK scheme, then $C_1^k$ can be found by Theorem \ref{sumrate}. This suggests that if $a_k\geq \sqrt{1+P_{k+1}}h_k$, user $k+1$ will be able to decode and hence remove the interference without affecting its own rate. Hence $C_1^k+C_{k+1}^K$ is achievable for the original $K$-user CGZIC.
\end{IEEEproof}

\end{document}